\documentclass[conference]{IEEEtran}
\usepackage{graphicx,color,psfrag}
\usepackage{amsmath}
\usepackage{psfrag}
\usepackage{subfigure}
\usepackage{color}
\usepackage{url}
\usepackage{epsfig}

\usepackage{ulem}
\usepackage[backend=bibtex,style=ieee]{biblatex}
\bibliography{Propagation,WSN,Control,Network,Books}
\AtEveryBibitem{\clearfield{url}}
\AtEveryBibitem{\clearfield{doi}}
\AtEveryBibitem{\clearfield{isbn}}
\AtEveryBibitem{\clearfield{issn}}

\newcommand{\Mx}[1]{{\textbf{#1} }}

\begin{document}

\title{Petroleum Refinery Multi-Antenna Propagation Measurements}

\author{Mohamed Gaafar and Geoffrey G.~Messier}

\maketitle

\begin{abstract}
This paper presents the results of the first multi-antenna propagation measurement campaign to be conducted at an operating petroleum refining facility.  The measurement equipment transmits pseudo-random noise test signals from two antennas at a 2.47~GHz carrier with a signal bandwidth of approximately 25~MHz.  The measurement data is analyzed to extract path loss exponent, shadowing distribution, fading distribution, coherence bandwidth and antenna correlation.  The results reveal an environment where large scale attenuation is relatively mild, fading is severe and good performance is expected from both antenna and frequency diversity.
\end{abstract}

\IEEEpeerreviewmaketitle

\section{Introduction}

Petroleum refining facilities are an important potential growth area for wireless technology.  Refining facilities have made extensive use of {\em wired} data networks for over 35 years \cite{zurawski-r-2014} and some refineries now have over 20,000 networked sensors and actuators \cite{young-re-2006}.  There is also clear economic incentive for petroleum operators to move to wireless links.  The metal safety shielding for wired cables in refineries can cost between \$200 and \$2,000 per foot \cite{johnstone-i-2007}.  Replacing these wired links with wireless would eliminate this cost.

\begin{figure}[htbp]
  \centerline{\includegraphics[width=2.75in]{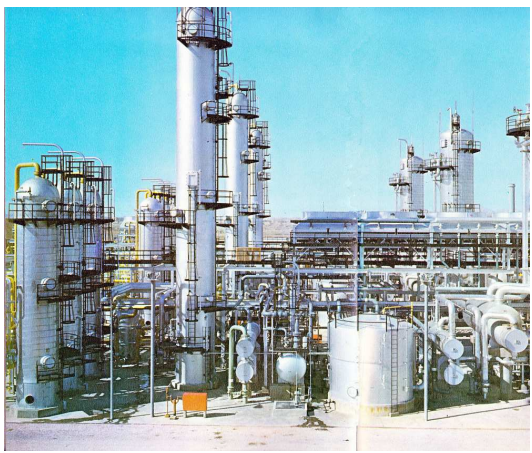}}
  \caption{Gas refinery.}
  \label{fg.env}
\end{figure}

One barrier to wireless adoption is our lack of understanding of wireless propagation within a refinery.  Wireless will mainly be used to connect outdoor sensors and actuators attached to pipes and vessels in an environment similar to Fig.~\ref{fg.env}.  This unique environment has arguably the greatest concentration of metallic scattering objects found anywhere.  To date, industrial propagation measurement studies have focused on the manufacturing and factory environment \cite{miranda-j-2013, coll-jf-2012, luo-s-2011, kozlowski-s-2008,  tanghe-e-2008, kunish-j-2006, miaoudakis-a-2005, rappaport-ts-1991, yegani-p-1991}.  However, the indoor factory environment is clearly very different than Fig.~\ref{fg.env}.   The only effort to date characterizing refinery propagation is \cite{savazzi-s-2013} which presents a modified path loss model  based on classifying the physical environment into different categories.  This model is verified using average received power measurements collected using wireless modems but the results are limited to path loss only. 

This paper presents the first multi-antenna propagation measurement campaign to be conducted at an operating gas refinery that captures both large and small scale channel statistics.  Sections~\ref{sec:setup} and \ref{sec:env} discuss the measurement equipment and environment, respectively.  Section~\ref{sec:analysis} presents the methods used to analyze the data.  Results are presented in Section~\ref{sec:results} with concluding remarks made in Section~\ref{sec:concl}.

\section{Equipment}
\label{sec:setup}

\begin{figure}[htbp]
  \centerline{\includegraphics[width=2.75in]{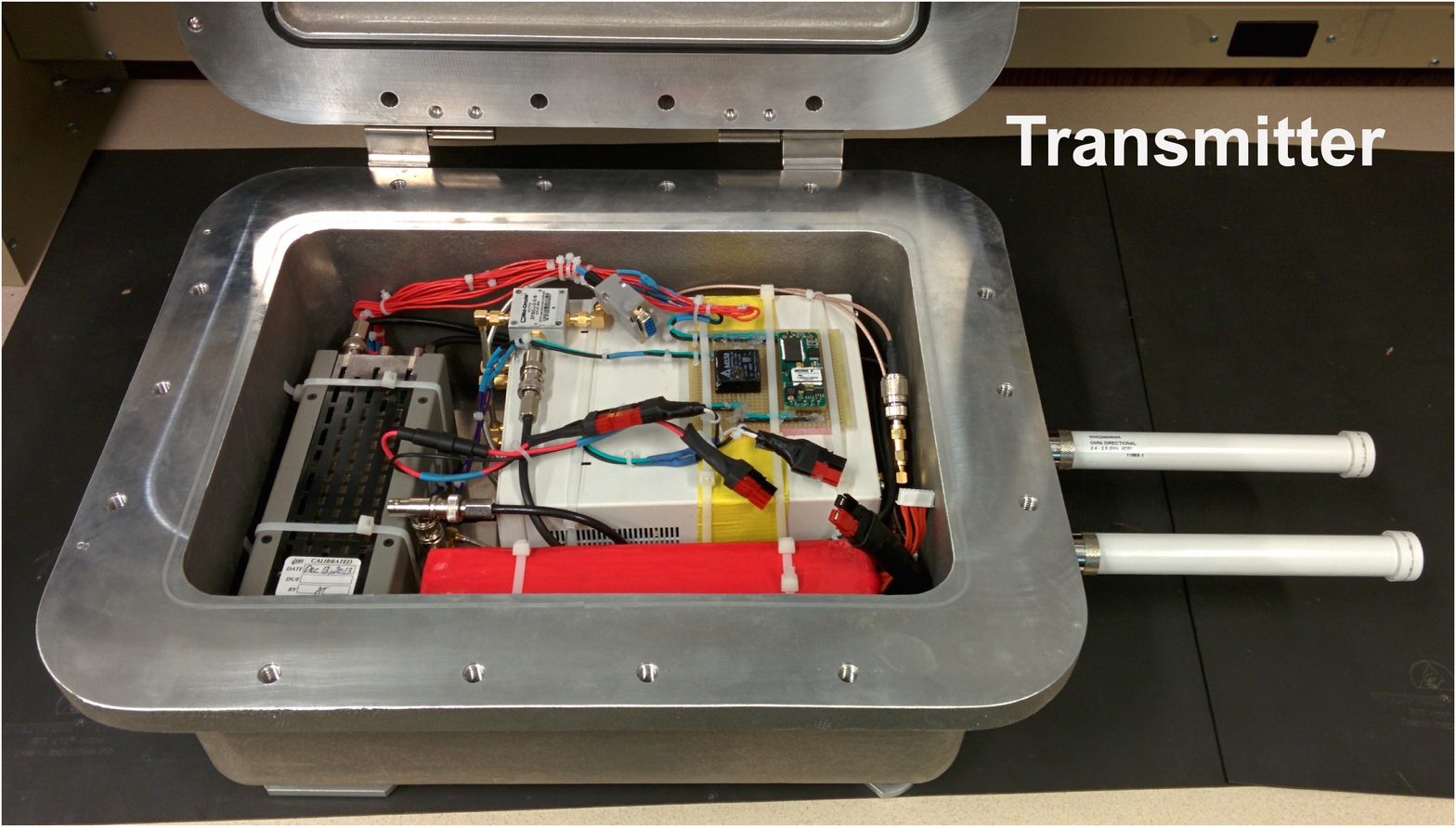}}
  \centerline{\includegraphics[width=2.75in]{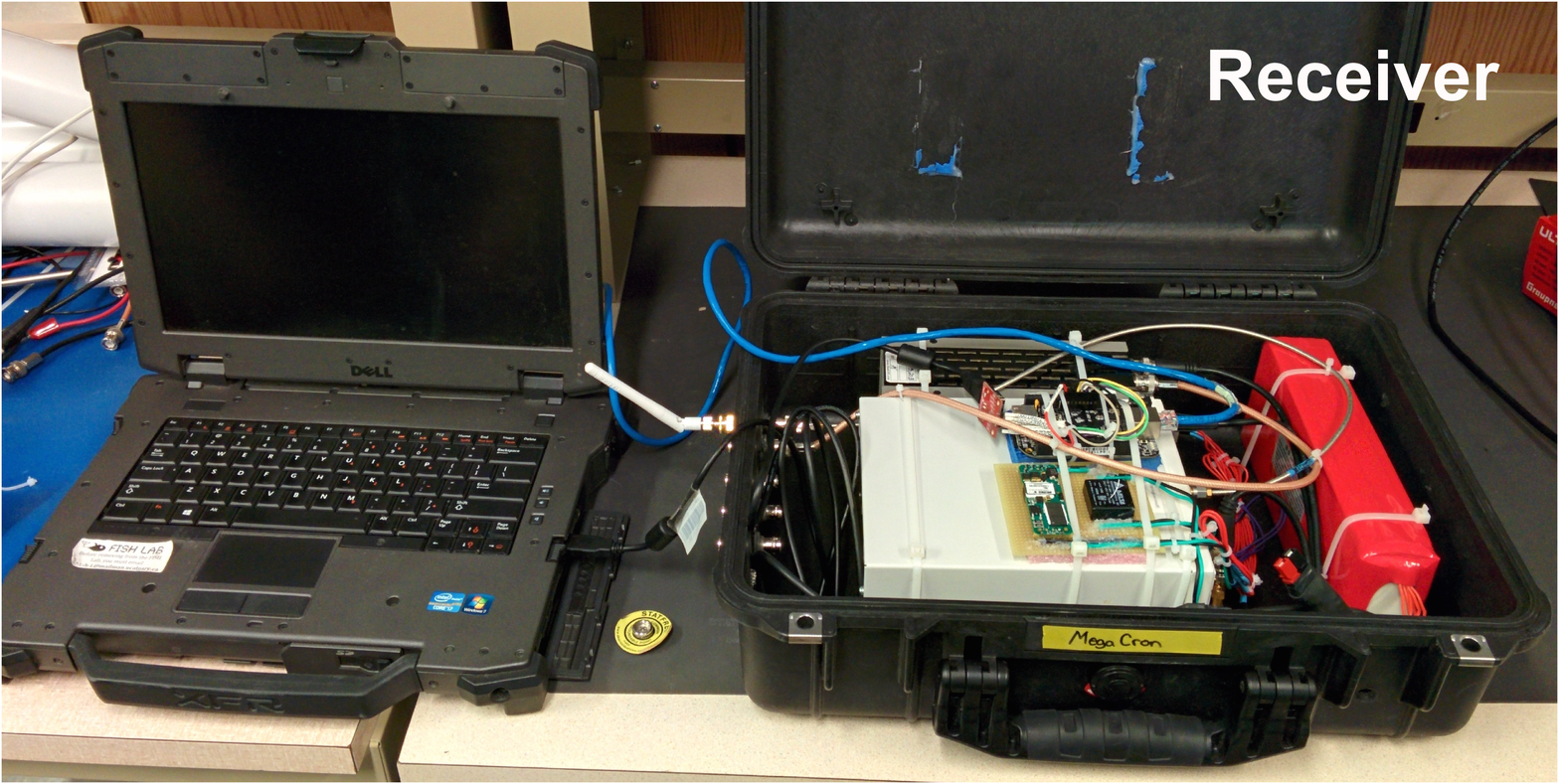}}
  \caption{Measurement equipment.}
  \label{fg.equip}
\end{figure}

The measurement transmitter, shown in Fig.~\ref{fg.equip}, transmits different time offsets of the same 25~Mchip/sec maximal length $2^{19}-1$ chip pseudo-random noise (PN) sequence \cite{proakis_jg1} from two PCTEL MHO24004NM dipole antennas \cite{mho24004nm}.  Antenna separation is approximately 16~cm.  The two offset sequences are denoted $p_1[n]$ and $p_2[n]$.  The signals are transmitted at 6~dBm equivalent isotropically radiated power (EIRP) at a carrier frequency of 2.4724~GHz.  The measurement receiver correlates the received signal with $p_1[n]$ and $p_2[n]$ in real time.  The correlation is performed over 100,000 chips so that a $2 \times 1$ MISO channel impulse response is captured on a laptop every 4~ms for a measurement capture rate of 250~Hz.  The transmitter and receiver are synchronized in time and frequency using SIM940 rubidium clock references \cite{sim940}.

Due to the potential of explosive vapors in the air, all electrical equipment used near petroleum refinery equipment must meet the Class 1/Division 1 (C1/D1)  standard which specifies all equipment must be shielded to prevent acting as an ignition source \cite{c1d1-2012}.  The transmit unit in Fig.~\ref{fg.equip} is contained in a C1/D1 compliant enclosure.

\section{Refinery Environment}
\label{sec:env}

\begin{figure}[htbp]
  \centerline{\includegraphics[width=3.5in]{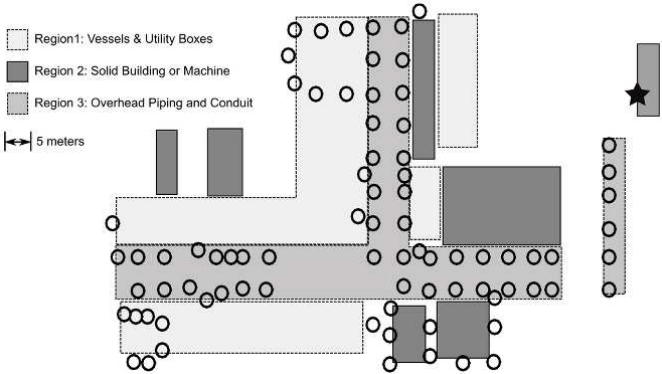}}
  \caption{Measurement environment.}
  \label{fg.map}
\end{figure}

The measurements were conducted at a Shell Canada gas refinery and a simplified map of the measurement area is shown in Fig.~\ref{fg.map}.  The environment consists of three regions: areas with tall vessels (10~m to 30~m in height) and piping similar to Fig.~\ref{fg.env}, solid metallic buildings or equipment that were opaque to radio signals and roadways covered by overhead piping and conduit approximately 2.5~m in the air.   The measurements characterize propagation conditions between 80 different candidate sensor/actuator locations, indicated by the black circles, and a central building housing the refinery process control hardware, indicated by the star.

At each sensor/actuator location, the C1/D1 compliant transmitter was held so the antennas were approximately 1.25~m from the ground.  During the measurement, the antennas were moved in a grid approximately 0.5~m $\times$ 0.5~m to capture local spatial variation.  The receiver was located outside the C1/D1 zone and connected to a single MHO24004NM dipole mounted 15~m in the air on the outside of the central process control building.

\section{Data Analysis}
\label{sec:analysis}

The impulse response of the channel between transmit antenna $i$ and the receiver is represented by matrix $\Mx{h}_i \in {\cal C}^{N \times L}$ where $N$ is the number of channel impulse responses captured and $L$ is the number of discrete channel taps.  The element at row $n$ and column $l$ of $\Mx{h}_i$ is denoted $h_i(n,l)$.

To determine large scale channel effects, an average attenuation is calculated for measurement location $r$ by averaging the signal received from antenna i according to $A_{r,i} = \frac{1}{N}\sum_n\sum_l |h_i(n,l)|^2$.  Unlike \cite{savazzi-s-2013}, large scale propagation is analyzed using a standard path loss and shadowing model. The path loss exponent is the slope of a best fit line on a log-log plot of attenuation versus distance and the shadowing values are the amount that each attenuation point deviates from that line \cite{rappaport-ts-1991}.

The power delay profile (PDP) of the channel is determined by averaging the power of $\Mx{h}_i$ along the $n$ axis.  The PDP is then used to calculate an RMS delay spread, $T_{rms}$, for each measurement as described in \cite{rappaport-ts-1991}.  The coherence bandwidth of the channel is then defined as $1/(5T_{rms})$ \cite{rappaport-ts-1991}.

To analyze small scale fading statistics, each measurement $\Mx{h}_i, i \in \{1,2\}$ for each location is normalized to have unit average power along the $n$ axis.  The channel responses are then transformed into the frequency domain by taking the discrete Fourier transform of each $h_i(n,l)$ along index $l$.  Let $H_i(n,w)$ denote the transformed impulse response where $w$ is the discrete frequency index and the definitions of $n$ and $i$ are unchanged.  The channel frequency response matrix is defined as $\Mx{H}_i \in {\cal C}^{N \times L}$, where $H_i(n,w)$ is the matrix element on the $n$th row and $w$th column.

The number of frequency response values with independent small scale fading is $R = \lfloor W_{\rm sig}/W_{\rm coh} \rfloor$, where  $W_{\rm coh}$ is the discrete coherence bandwidth of the channel and $W_{\rm sig}$ is the discrete bandwidth of the measurement signal.  Since $N$ fading values are captured for each of these frequency points, this allows the creation of  a $1 \times NR$ {\em fading vector} for the $i$th transmit antenna defined as $\Mx{e}_i = [\ |H_i(0,0)|\ \ldots\ |H_i(N-1,0)|\ |H_i(0,W_{\rm coh})| \ldots\ |H_i(N-1,RW_{\rm coh})|\ ]$.

Small scale fading severity is quantified using Ricean K-factor.  Before K-factor is estimated, the Chi-square goodness-of-fit test \cite{navidi_wc1} is applied to the fading vector to ensure its envelope variations can be accurate represented by the Ricean distribution.  A significance level of 10\% is used, meaning there is a 10\% chance that measurements that are actually Ricean will be discarded.  K-factor is estimated for all measurements satisfying the Chi-square test using the method of moments \cite{greenstein-lj-1999}.  Antenna correlation is calculated by determining the correlation coefficient between $\Mx{e}_1$ and $\Mx{e}_2$.  A correlation value is calculated for all measurements, regardless of whether they satisfy the Ricean fit test.

\section{Results}
\label{sec:results}

\begin{figure}[htbp]
  \centerline{\includegraphics[width=3in]{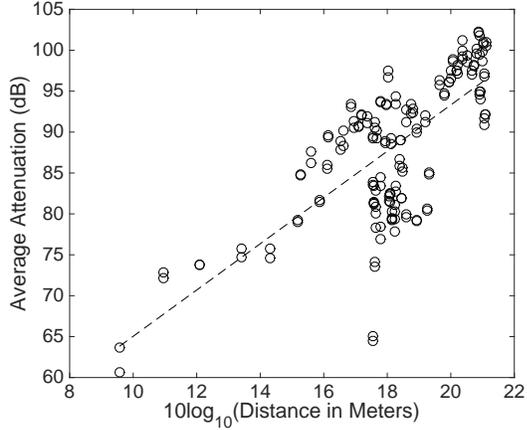}}
  \caption{Log average attenuation versus log distance.}
  \label{fg.loglog}
\end{figure}

\begin{figure}[htbp]
  \centerline{\includegraphics[width=3in]{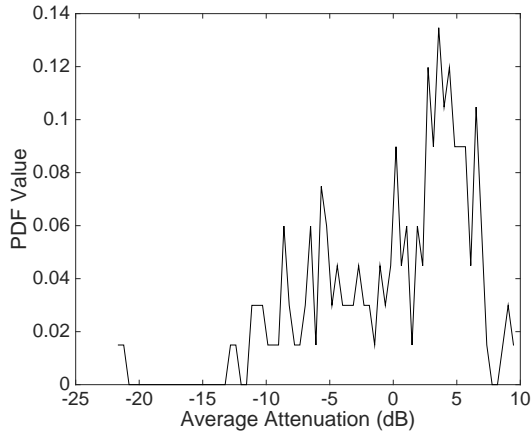}}
  \caption{Shadowing distribution.}
  \label{fg.shadow}
\end{figure}

The results of the large scale analysis are shown in Figs.~\ref{fg.loglog} and \ref{fg.shadow}.  The slope of the best fit line in Fig.~\ref{fg.loglog} yields a path loss exponent of 2.76 and shadowing standard deviation is 5.93~dB.  While both the shadowing and path loss values seem low given the amount of clutter, it is important to remember the highly metallic nature of the environment.  Since the majority of the scattering objects are metallic, reflection is the dominant propagation mechanism with less absorption and diffusion than occurs in other environments.  On average, this results in a higher proportion of the transmit power reaching the receiver.

It is also clear from Fig.~\ref{fg.shadow} that shadowing does not display the traditional log-normal distribution.  A true log-normal distribution requires the power of the received signal to be the product of many random attenuation factors due to scatterers along the propagation path \cite{suzuki-h-1977}.  We note that a refinery has at least as many objects contributing to shadowing as the indoor wireless channel and that indoor shadowing is much closer to log-normal \cite{hong-a-2006}.  Therefore, we assume that the cause of the unusual shadowing distribution is not a lack of scattering events.  Instead, the mostly metallic environment does not create a sufficiently random change in amplitude and phase during each scattering event for the product of those events to quickly approach a log-normal distribution.

\begin{figure}[htbp]
  \centerline{\includegraphics[width=3in]{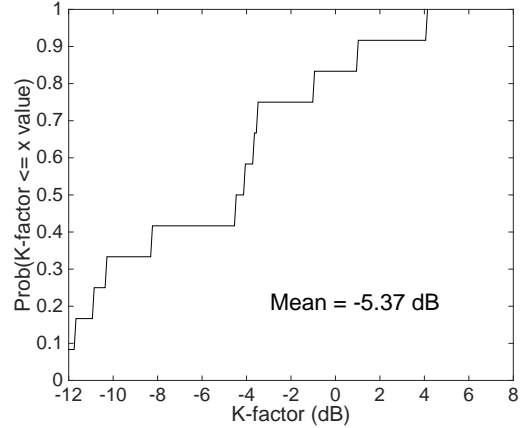}}
  \caption{CDF of K-factor.}
  \label{fg.kfactor}
\end{figure}

The CDF of the measurement K-factor values shown in Fig.~\ref{fg.kfactor} indicates very severe fading.  With a mean K-factor of -5.37~dB, the fading is approximately Rayleigh.  This is not surprising given the dense number of scattering objects surrounding the locations in Fig.~\ref{fg.map}.  However, since the vast majority of refinery sensors and actuators are fixed to one location and none of the scattering objects move, Fig.~\ref{fg.kfactor} underscores that utilizing diversity techniques will be extremely important since a sensor in a deep fade cannot rely its motion or motion in the environment to improve its propagation conditions.

\begin{figure}[htbp]
  \centerline{\includegraphics[width=3in]{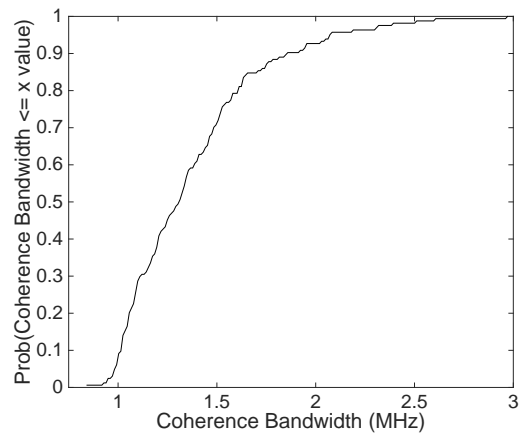}}
  \caption{CDF of coherence bandwidth.}
  \label{fg.cohBw}
\end{figure}

\begin{figure}[htbp]
  \centerline{\includegraphics[width=3in]{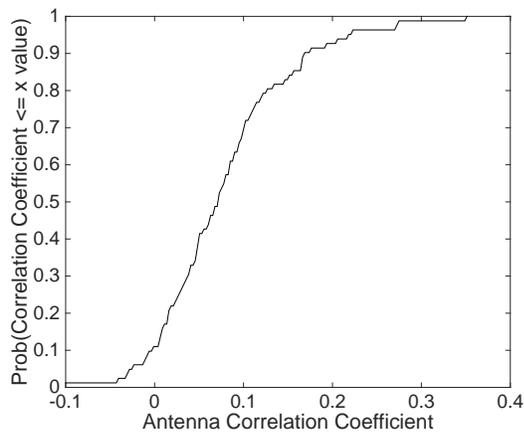}}
  \caption{CDF of antenna correlation coefficient.}
  \label{fg.cc}
\end{figure}

Fortunately, the measurements indicate that both antenna diversity and frequency diversity through channel hopping will be effective.  Fig.~\ref{fg.cohBw} shows the CDF of the coherence bandwidth of the measured channels.  A mean of 1.38~MHz corresponds to a mean RMS delay spread of 154~ns.  ISA 100.11a is a representative industrial wireless network standard \cite{isa100.11a} that utilizes an 802.15.4 physical layer and makes extensive use of channel hopping.  As the channels for this standard are separated by 5~MHz, this coherence bandwidth suggests that the channels will enjoy the independent fading necessary to make frequency hopping an effective fading mitigation option.  Fig.~\ref{fg.cc} shows the CDF of the antenna fading correlation coefficient is very low with a mean value of 0.079.  This means that multiple antennas will also be effective for those nodes where size and cost allow for them.

\section{Conclusion}
\label{sec:concl}

This paper has presented the first multi-antenna measurement campaign conducted at an operating petroleum refinery that captures both large and small scale channel statistics.  The results reveal an environment where path loss and shadowing are milder than the number of obstacles would suggest due to the large number of reflecting metallic objects that are present.  Fading is severe due to the large number of scatterers but low values for both coherence bandwidth and antenna correlation coefficient indicate that frequency and antenna diversity are both good options for mitigating this fading.


\printbibliography

\end{document}